\documentclass[twocolumn,pra]{revtex4}
\usepackage{amsmath}
\usepackage{amstext}
\usepackage{latexsym}
\usepackage{psfig}
\usepackage{graphicx}
\usepackage{amsfonts}

\newcommand{\ket}[1]{\left\vert#1\right\rangle}

\begin{document}
  
  \title{Schr\"odinger Cat States for Quantum Information Processing
}
  
\author{H. Jeong and T.C. Ralph}
  
\affiliation{
Department of Physics, University of Queensland, St Lucia, Qld 4072,
   Australia }
  
  \date{\today}  

\begin{abstract}
 We extensively discuss how 
Schr\"odinger cat states (superpositions
of well-separated coherent states) in optical systems can
be used for quantum information processing.
\end{abstract}

  \maketitle




\section{Introduction}

In the early days of quantum mechanics many of its founders became very worried by some of the 
paradoxical predictions that emerged from thought experiments based on the new theory. Now, eighty 
years on, some of these early thought experiments are being experimentally realized, and more than just 
confirming the fundamentals of the theory they are also being recognized as the basis of 21st century 
technologies \cite{Dow}. An example is the EPR paradox, proposed by Einstein, Podolsky and Rosen in 
1935 \cite{Ein}, which discussed the strange properties of quantum entanglement. Today, entanglement 
has been observed in optical \cite{Asp,Kwi} and ion \cite{Sac} systems and is recognized as a resource 
for many quantum information processing tasks \cite{Nie}. 

About the same time as the EPR discussion, Schr{\"o}dinger proposed his famous cat paradox \cite{Schr1} 
that highlighted the unusual consequences of extending the concept of superposition to macroscopically 
distinguishable objects. From a quantum optics view point, the usual paradigm is to consider 
superpositions of coherent states with amplitudes sufficiently different that they can be resolved 
using homodyne detection \cite{Leg,Rei}. In this chapter we discuss how, beyond their fundamental 
interest, these types of states can be used in quantum information processing. We then look at the 
problem of producing such states with the required properties

\section{Quantum Information Processing 
with Schr\"odinger Cat States}

\subsection{Coherent-state qubits}

We now introduce qubit systems using coherent states.
A coherent state can be defined as \cite{Schr2,Cahill}
\begin{equation}
|\alpha\rangle=
e^{-|\alpha|^2/2}\sum_{n=0}^{\infty}\frac{\alpha^n}{\sqrt{n!}}|n\rangle,
\end{equation}
where $|n\rangle$ is a number state and
$\alpha$ is the complex amplitude of the coherent state.
The coherent state is a very 
useful tool in quantum optics and a laser field is considered a good
approximation of it.
Let us consider two coherent states $|\alpha\rangle$ and
$|-\alpha\rangle$.
 The two coherent states are not orthogonal to each other but
their overlap $|\langle\alpha|-\alpha\rangle|^2=e^{-4|\alpha|^2}$
decreases exponentially with $|\alpha|$.  For example, when $|\alpha|$ is
as small as 2, the overlap is $\approx 10^{-7}$,
{\it i.e.}, $|\langle\alpha|-\alpha\rangle|^2\approx0$.
  We identify the two coherent states of
$\pm\alpha$ as basis states for a logical qubit as
$|\alpha\rangle\rightarrow|0_L\rangle$ and
$|-\alpha\rangle\rightarrow|1_L\rangle$,
so that a qubit state is represented by 
\begin{equation}
\label{c-qubit}
|\phi\rangle={\cal  A}|0_L\rangle+{\cal B}|1_L\rangle={\cal  A}|\alpha\rangle+{\cal B}|-\alpha\rangle.
\end{equation}
The basis states,
$|\alpha\rangle$ and $|-\alpha\rangle$, can be
unambiguously
 discriminated by a
simple measurement scheme with a 50-50 beam splitter, an auxiliary
coherent field of amplitude $\alpha$ and two photodetectors 
\cite{JK}. At the beam splitter, the qubit state
$|\phi\rangle_1$ is mixed with the auxiliary state
$|\alpha\rangle_2$ and results in the output
\begin{equation}
\label{eq:ax}
|\phi_R\rangle_{ab}={\cal A}|\sqrt{2}\alpha\rangle_a|0\rangle_b+{\cal 
B}|0\rangle_a|-\sqrt{2}\alpha\rangle_b.
\end{equation}
The two photodetector are set for modes $a$ and $b$ respectively.  
If detector $A$ registers any photon(s) while detector $B$
does not, we know that
$|\alpha\rangle$ was measured. On the contrary, if $A$ does not click
while $B$ does, the
measurement outcome was $|-\alpha\rangle$.  Even though there is
non-zero probability of failure 
$P_f(\phi_R)=|\langle00|\phi_R\rangle|^2=|{\cal
  A}+{\cal   B}|^2\mbox{e}^{-2\alpha^2}$
 in which both of the detectors
do not register a photon, the failure is known from the result whenever
it occurs, and $P_f$ approaches to zero exponentially as $\alpha$
increases. 
Note that the detectors do {\it not} have to be highly efficient 
for unambiguous discrimination.
Alternatively, homodyne detection can also be very efficient
for the qubit readout because 
the overlap between the coherent states $|\alpha\rangle$
and $|-\alpha\rangle$ would be extremely small for an appropriate
value of $\alpha$.

Alternatively, it is possible to construct an
exactly
 orthogonal qubit basis
with 
the equal superposition of 
two linear independent coherent
states $|\alpha\rangle$ and $|-\alpha\rangle$. Consider the basis states
\begin{eqnarray}
&&|e\rangle={\cal N}_+(|\alpha\rangle+|-\alpha\rangle)
\rightarrow|0_L\rangle,\\
&&|d\rangle={\cal N}_-(|\alpha\rangle-|-\alpha\rangle)\rightarrow|1_L\rangle,
\end{eqnarray}
where 
${\cal N}_\pm=1/\sqrt{2(1\pm\exp[-2|\alpha|^2])}$.
 It can be
simply shown that they form an orthonormal basis as
$\langle e |d\rangle= \langle d |e\rangle=0$ and 
$\langle e | e\rangle=\langle d| d\rangle=1$.
The basis state $|e\rangle$ ($|d\rangle$) is called ``even cat state''
(``odd cat state'')
because it contains
only even (odd) number of photons as
\begin{eqnarray}
|e\rangle 
&=&2{\cal N}_+e^{-\frac{|\alpha|^2}{2}}\sum_{n=0}^\infty\frac
{\alpha^{2n}}{\sqrt{(2n)!}}|2n\rangle,\\
|d\rangle
&=&2 {\cal N}_-e^{-\frac{|\alpha|^2}{2}}\sum_{n=0}^\infty
\frac{\alpha^{(2n+1)}}{\sqrt{(2n+1)!}}|2n+1\rangle.
\end{eqnarray}
The even and odd cat states can thus be discriminated
by a photon parity measurement which can be represented by 
$O_\Pi=\sum_{n=0}^\infty(|2n\rangle\langle 2n|-|2n+1\rangle\langle2n+1|)$.
As  $\alpha$ goes to zero, the odd cat state $|d\rangle$
 approaches a single photon state $|1\rangle$  while 
the even cat state $|e\rangle$ approaches $|0\rangle$.
No matter how small $\alpha$ is, there is no possibility that no photon
will be detected from the state $|d\rangle$ at an ideal photodetector.

\subsection{Quantum teleportation}

Quantum teleportation is an interesting phenomenon
for demonstrating 
quantum theory 
and a useful tool in quantum information processing \cite{Bennett93}.
By quantum teleportation, an unknown
quantum state is disentangled in a sending place and its perfect replica
appears at a distant place via dual quantum and classical channels.
The key ingredients of quantum teleportation are 
an entangled channel,
a Bell-state
measurement and appropriate unitary transformations.
In what follows we shall explain how teleporation
can be performed for a coherent-state qubit \cite{Enk,JKL01}.   

Let us assume that Alice wants to teleport an unknown coherent-state qubit
$|\phi\rangle_a$ 
via a pure entangled coherent channel
\begin{equation}
|\Psi_-\rangle_{bc}=N_-(|\alpha\rangle_b|
-\alpha\rangle_c-|-\alpha\rangle_b|\alpha\rangle_c),
\end{equation}
where $N_-$ is the normalization factor.
 After
sharing the quantum channel $|\Psi_-\rangle$,
 Alice should perform a
Bell-state measurement on her part of the quantum channel and the
unknown qubit $|\phi\rangle$  and send the outcome to Bob.
The Bell-state measurement is to discriminate between 
the four Bell-cat states 
 which can be defined with coherent states as
 \cite{Sanders,Sanders95,Hirota01,Hirota01b}
\begin{eqnarray}
\label{eq:qbs1}
&&|\Phi_\pm\rangle=N_\pm(|\alpha\rangle|\alpha\rangle
\pm|-\alpha\rangle|-\alpha\rangle),\\
\label{eq:qbs2}
&&|\Psi_\pm\rangle=N_\pm(|\alpha\rangle|
-\alpha\rangle\pm|-\alpha\rangle|\alpha\rangle),
\end{eqnarray}
where $N_\pm$ are normalization factors.
The four Bell-cat
states defined in our framework
are a very good approximation of the
Bell basis. 
These states are orthogonal to each other except
$\langle\Psi_+|\Phi_+\rangle=1/\cosh 2|\alpha|^2$,
and 
$|\Psi_+\rangle$ and $|\Phi_+\rangle$ rapidly become
orthogonal as $|\alpha|$ grows.

A Bell-state measurement, or simply Bell measurement, is very useful in
quantum information
processing. 
It was shown that a complete Bell-state measurement on a product
Hilbert space of two two-level systems is not possible using linear
elements \cite{L}.
A Bell measurement scheme using linear optical elements \cite{Br95} has
been used  to distinguish 
 only up to  two of the Bell states for
teleportation \cite{Bou97exp} and dense coding \cite{Mattle96}.
 However, a remarkable feature of the Bell-cat
states is that each one of them can be unambiguously discriminated using only
a beam splitter and photon-parity measurements \cite{JKL01,JKpuri}.
Suppose that the modes, $a$ and $b$, of the entangled state are
incident on a 50-50 beam splitter. After passing the beam splitter, the
Bell-cat states become
\begin{eqnarray}
\label{fig:setup1}
|\Phi_+\rangle_{ab} &\longrightarrow& 
  |E\rangle_f|0\rangle_g,
  \nonumber \\
  |\Phi_-\rangle_{ab} &\longrightarrow &
  |D\rangle_f|0\rangle_g,
  \nonumber \\
  |\Psi_+\rangle_{ab} &\longrightarrow &
  |0\rangle_f|E\rangle_g, 
  \nonumber \\
  |\Psi_-\rangle_{ab} &\longrightarrow &
  |0\rangle_f|D\rangle_g,
\end{eqnarray}
where the even cat state $|E\rangle\propto|\sqrt{2}\alpha\rangle
+|-\sqrt{2}\alpha\rangle$ 
definitely contains an even number of photons, while the odd cat state
$|D\rangle\propto|\sqrt{2}\alpha\rangle
-|-\sqrt{2}\alpha\rangle$
definitely contains an odd number of photons.
  By setting two
photodetectors for the output modes $f$ and $g$ respectively to perform number
parity measurement, 
the Bell-cat measurement can be simply achieved.  For example, if
an odd number of photons is detected for mode $f$, the state
$|\Phi_-\rangle$ is measured, and if an odd number of photons is
detected for mode $g$, then $|\Psi_-\rangle$ is measured.  Even though
there is non-zero probability of failure in which both of the
detectors do not register a photon due to the non-zero overlap of
$|\langle0|E\rangle|^2=2e^{-2|\alpha|^2}/(1+e^{-4|\alpha|^2})$,
it is small for an appropriate choice of
$\alpha$ and the failure is known from the result whenever it occurs.

To complete the teleportation process,
Bob 
performs a unitary transformation on his part of the quantum channel
according to the measurement result sent from Alice
via a classical channel.
The required transformations are $\sigma_x$ and $\sigma_z$ 
on the coherent-state qubit basis, where
$\sigma$'s are Pauli operators.
When the measurement outcome is $|B_4\rangle$, 
Bob obtains a perfect replica of the 
original unknown qubit without any
operation.
When the measurement outcome is $|B_2\rangle$, Bob should perform
$|\alpha\rangle\leftrightarrow|-\alpha\rangle$ on his qubit. Such a phase
shift by $\pi$ can be done using a phase shifter whose action is
described by $P(\varphi)=\mbox{e}^{i\varphi a^\dag a}$,
where $a$ and $a^\dagger$ are the annihilation and creation operators.
When the outcome is $|B_3\rangle$, 
the transformation should be performed as 
$|\alpha\rangle\rightarrow|\alpha\rangle$ and
$|-\alpha\rangle\rightarrow-|-\alpha\rangle$.
This transformation is more difficult but can be achieved 
most straightforwardly by simply teleporting the state again (locally) and
repeating until the required phase shift is obtained.
Therefore, both of the required unitary transformation, $\sigma_x$ and $\sigma_z$,
 can be performed by
linear optics elements.
 When the outcome is $|B_1\rangle$, $\sigma_x$ and
$\sigma_z$ should be successively applied.

\subsection{Quantum computation}

We now describe how a universal set of quantum gates can be implemented on coherent state qubits using 
only linear optics and photon detection, provided a supply of cat states is available as a resource. 
The idea was originally due to Ralph, Munro and Milburn \cite{Ralph} and was later expanded on by Ralph 
{\it et al} \cite{Ralph2}.

\begin{figure}
\centerline{\scalebox{.5}{\includegraphics{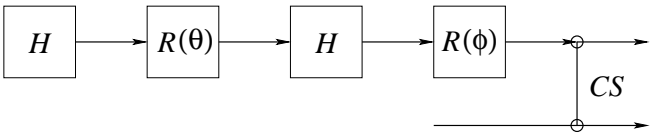}}} 
\caption{
A set of Hadamard ($H$) gates, rotations ($R$) about the Z-axis and control sign ($CS$) gates can 
provide
universal gate operations.} 
\label{fig:gate}
\end{figure}

A universal single qubit quantum gate element can be constructed from the following sequence of gates: 
Hadamard ($H$); rotation about the Z-axis by angle $\theta$ $(R(\theta))$; Hadamard ($H$) and; rotation 
about the Z-axis by angle $\phi$ $(R(\phi))$. If the two qubit gate, control sign ($CS$), is also 
available then universal processing is possible (See Fig.~\ref{fig:gate}).
We now describe how these gates can be implemented.
We will assume that deterministic single qubit measurements can be made in the computational basis, 
$\ket{\alpha}, \ket{-\alpha}$ and the phase superposition basis $\ket{\alpha} \pm \exp[i \epsilon] 
\ket{-\alpha}$. As described in the previous section, computational basis measurements can be achieved 
using either homodyne or photon counting techniques. The phase superposition basis can be measured 
using photon counting in a Dolinar receiver type arrangement \cite{Dol,Tak}. The simplest case is for 
$\epsilon = 0$ where we need to differentiate only between odd or even photon numbers in direct 
detection. We also assume we can make two qubit Bell-measurements and, more generally, perform 
teleportation, as described in the previous section.

{\it Hadamard Gate}: The Hadamard gate ($H$) can be defined by its effect on the computational states: 
$H \ket{\alpha} = \ket{\alpha} + \ket{-\alpha}$ and $H \ket{-\alpha} = \ket{\alpha} - \ket{-\alpha}$ 
where for convenience we have dropped normalization factors. One way to achieve this gate is to use the 
resource state $\ket{HR} = \ket{\alpha, \alpha} + \ket{\alpha, -\alpha} + \ket{-\alpha, \alpha} - 
\ket{-\alpha, -\alpha}$. This state can be produced non-deterministically from cat state resources, as 
will be described shortly. It is straight forward to show that if a Bell-state measurement is made 
between an arbitrary qubit state $\ket{\sigma}$ and one of the modes of $\ket{HR}$ then the remaining 
mode is projected into the state $H \ket{\sigma}$, where dependent on the outcome of the 
Bell-measurement a bit-flip correction, a phase-flip correction, or both may be necessary. 

{\it Phase Rotation Gate}: The phase rotation gate $(R(\theta))$ can be defined by its effect on the 
computational states: $R(\theta) \ket{\alpha} = \exp[i \theta] \ket{\alpha}$ and $R(\theta) 
\ket{-\alpha} = \exp[-i \theta] \ket{-\alpha}$. One way to achieve this gate is the following: The 
arbitrary qubit, $\mu \ket{\alpha} + \nu \ket{-\alpha}$ is split on a 50:50 beamsplitter giving the two 
mode state: $\mu \ket{\alpha/\sqrt{2}}\ket{\alpha/\sqrt{2}} + \nu 
\ket{-\alpha/\sqrt{2}}\ket{-\alpha/\sqrt{2}}$. One of the modes is then measured in the phase 
superposition basis $\ket{\alpha/\sqrt{2}} \pm \exp[-2i \theta] \ket{-\alpha/\sqrt{2}}$, thus 
projecting the other mode into the state $\mu \exp[i \theta] \ket{\alpha/\sqrt{2}} \pm \nu \exp[-i 
\theta] \ket{-\alpha/\sqrt{2}}$. The amplitude decrease can be corrected by teleportation in the 
following way \cite{Ralph2}. The asymmetric Bell state entanglement, $\ket{\alpha/\sqrt{2}}\ket{\alpha} 
+ \ket{-\alpha/\sqrt{2}}\ket{-\alpha}$ is produced by splitting the cat state $\ket{\sqrt{3/2}\alpha} + 
\ket{-\sqrt{3/2}\alpha}$ on a $1/3:2/3$ beamsplitter. Teleportation is then carried out with the Bell 
state measurement being performed between the matching ``$\alpha/\sqrt{2}$'' modes and the teleported 
state ending up on the ``$\alpha$'' mode. Dependent on the outcome of the phase basis measurement and 
the Bell-measurement a bit-flip correction, a phase-flip correction, or both may be necessary.

{\it Control Sign Gate}: The control-sign gate $(CS)$ can be defined by its effect on the two qubit 
computational states: $CS \ket{\alpha}\ket{\alpha} = \ket{\alpha}\ket{\alpha}$; $CS 
\ket{\alpha}\ket{-\alpha} = \ket{\alpha}\ket{-\alpha}$; $CS \ket{-\alpha}\ket{\alpha} = 
\ket{-\alpha}\ket{\alpha}$ and; $CS \ket{-\alpha}\ket{-\alpha} = -\ket{-\alpha}\ket{-\alpha}$. One way 
to achieve this gate is the following: The two arbitrary qubits, $\mu \ket{\alpha} + \nu \ket{-\alpha}$ 
and $\gamma \ket{\alpha} + \delta \ket{-\alpha}$ are both split on 50:50 beamsplitters giving the two 
mode states: $\mu \ket{\alpha/\sqrt{2}}\ket{\alpha/\sqrt{2}} + \nu 
\ket{-\alpha/\sqrt{2}}\ket{-\alpha/\sqrt{2}}$ and $\gamma \ket{\alpha/\sqrt{2}}\ket{\alpha/\sqrt{2}} + 
\delta \ket{-\alpha/\sqrt{2}}\ket{-\alpha/\sqrt{2}}$. A Hadamard gate is then performed on the second 
mode of the first qubit giving the state $\mu 
\ket{\alpha/\sqrt{2}}(\ket{\alpha/\sqrt{2}}+\ket{-\alpha/\sqrt{2}}) + \nu 
\ket{-\alpha/\sqrt{2}}(\ket{\alpha/\sqrt{2}}-\ket{-\alpha/\sqrt{2}})$. If a Bell-measurement is then 
carried out between the second mode of the first qubit and one of the modes of the second qubit a CS 
gate will be achieved. The amplitude reduction can be corrected as before using teleportation. 
Dependent on the outcome of the various Bell-measurements, bit-flip corrections, phase-flip 
corrections, or both may be necessary.

{\it Resource State}: The resource state  $\ket{HR}$ can be produced in the following way. Consider the 
beamsplitter interaction given by the unitary transformation
\begin{equation}
U_{ab}=\exp[i {{\theta}\over{2}} (a b^{\dagger}+a^{\dagger} b)]
\end{equation}
where $a$ and $b$ are the annihilation operators corresponding to two
coherent state qubits $|\gamma \rangle_{a}$ and $|\beta
\rangle_{b}$, with $\gamma$ and $\beta$ taking values of $-\alpha$ or
$\alpha$. It is well known that the output state
produced by such an interaction is
\begin{equation}
U_{ab} |\gamma \rangle_{a} |\beta \rangle_{b}=|\cos
{{\theta}\over{2}} \gamma+i
\sin {{\theta}\over{2}} \beta \rangle_{a} |\cos {{\theta}\over{2}}
\beta+
i \sin {{\theta}\over{2}} \gamma \rangle_{b}
   \label{Ho}
\end{equation}
where $\cos^{2} {{\theta}\over{2}}$ ($\sin^{2} {{\theta}\over{2}}$)
is the reflectivity
(transmissivity) of the beamsplitter. Suppose two cat states are fed into the beamsplitter and both 
output beams are then teleported, the output state will be:
\begin{eqnarray}
& &  e^{-\theta^{2} \alpha^{2}/4}
   (e^{i \theta \alpha^{2}}|-\alpha \rangle_{a} |-\alpha
   \rangle_{b}\pm
e^{-i \theta \alpha^{2}}|\alpha \rangle_{a} |-\alpha \rangle_{b}
\pm \nonumber\\
& & \; \; \; \; \; \; \; \; \; \; \; e^{-i \theta
\alpha^{2}}|-\alpha
\rangle_{a} |\alpha \rangle_{b}+
e^{i \theta \alpha^{2}}
|\alpha \rangle_{a} |\alpha \rangle_{b})
   \label{Ho1}
\end{eqnarray}
where the $\pm$ signs depend on the outcome of the
Bell
measurements. If we choose $\phi=2 \theta \alpha^{2}=\pi/2$ then
the resulting state is easily shown to be
locally equivalent to $\ket{HR}$ (related by phase rotations). Preparation of this state is 
non-deterministic because of non-unit overlap between the state of Eq. (\ref{Ho}) and the Bell states 
used in the teleporter. As a result the teleporter can fail by recording photons at both outputs in the 
Bell-measurement. The probability of success is $e^{-\theta^{2} \alpha^{2}/2}$. For $\alpha = 2$ this 
is about 92\% probability of success.

{\it Correction of Phase-flips}: After each gate we have noted that bit flip and/or phase flip 
corrections may be necessary since our gate operations are based on
the teleportation protocol. 
As discussed in the previous section, bit flips can be easily implemented
using a phase shifter, $P(\pi)$,
while phase-flips are more expensive.
We now argue that in fact only active correction of bit-flips is necessary. This is because phase-flips 
commute with the phase rotation gate and the control sign gate but are converted into bit flips by the 
Hadamard gate. This suggests the following strategy: After each gate operation any bit-flips are 
corrected whilst phase-flips are noted. After the next Hadamard gate the phase flips are converted to 
bit-flips which are then corrected and any new phase-flips are noted. By following this strategy only 
bit-flips need to be corrected actively, with, at worst, some final phase-flips needing to be corrected 
in the final step of the circuit.

\subsection{Entanglement purification for Bell-cat states}

It is not possible to perfectly isolate a quantum state
from its environment. A
quantum state inevitably loses its quantum coherence in a 
dissipative environment. This process is called decoherence 
and has been known as the main obstacle to 
the physical implementation of quantum information processing.
Quantum error correction \cite{Cochrane,Ralph2,Glancy}
and entanglement purification \cite{JKpuri,Clausen00} 
have been studied 
for quantum information processing using cat states
to overcome this problem. Here we discuss an entanglement
purification technique.

\begin{figure}
\centerline{\scalebox{.4}{\includegraphics{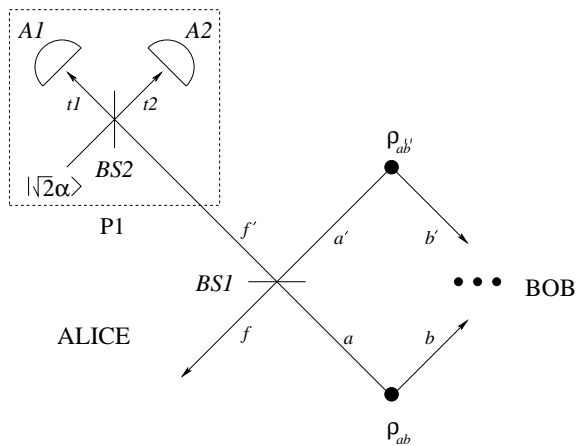}}} 
\caption{A schematic of the entanglement purification scheme for
     mixed entangled coherent states. P1 tests if the incident fields
     $a$ and $a^\prime$ were in the same state by simultaneous clicks
     at $A1$ and $A2$.  
	 } 
\label{fig:ep} 
\end{figure}

An entanglement purification for entangled coherent
states (Bell-cat states) have been studied by several authors 
\cite{JKpuri,Clausen00}.
It has been found that certain types of mixed states
including the Werner-type mixed states composed of the Bell-cat states
can be purified by 
simple linear optics elements and inefficient detectors
\cite{JKpuri}.
The other types of mixed states need to be transformed to the Werner
type states by local operations. 
This scheme performs amplification of the Bell-cat states
simultaneously with entanglement purification. 
This is an important observation because 
Bell-cat states of large amplitudes are preferred for quantum information
processing while
their generation is hard.
 A similar technique is employed to 
generate single-mode large cat states \cite{Lund04}.

We first explain the purification-amplification protocol 
for entangled coherent states 
by a simple example
and then apply it to a realistic situation 
\cite{JKpuri}.
Let us suppose that Alice and Bob want to distill entangled
coherent states $|\Phi_+\rangle$
from a type of ensemble
\begin{equation}
\label{ensemble}
\rho_{ab}=F|\Phi_+\rangle\langle\Phi_+|+G|\Psi_+\rangle\langle\Psi_+|,
\end{equation}
where $F+G\approx1$ for $|\alpha|\gg1$. 
We shall 
assume this condition, $|\alpha|\gg1$, for simplicity.
The purification-amplification process 
can be simply accomplished by performing
the process shown in Fig.~\ref{fig:ep}.
Alice and Bob choose two pairs from the ensemble which are
represented by the following density operator 
\begin{widetext}
\begin{eqnarray}
\label{eq:rho*2}
&&\rho_{a b} \rho_{a^\prime b^\prime}=F^2|\Phi_
+\rangle\langle\Phi_+|\otimes|\Phi_+\rangle\langle\Phi_+|
+F(1-F)|\Phi_+\rangle\langle\Phi_+|\otimes
|\Psi_+\rangle\langle\Psi_+|\nonumber\\
&&~~~~~~+F(1-F)|\Psi_+\rangle\langle\Psi_+|\otimes
|\Phi_+\rangle\langle\Phi_+|+(1-F)^2|\Psi_+\rangle\langle
\Psi_+|\otimes|\Psi_+\rangle\langle\Psi_+|.\nonumber\\
\end{eqnarray}
\end{widetext}
The fields of modes $a$ and $a^\prime$ are in Alice's possession while
$b$ and $b^\prime$ in Bob's. In Fig.~\ref{fig:ep}(a), we show that
Alice's action to purify the mixed entangled state. The same is
conducted by Bob on his fields of $b$ and $b^\prime$.

There  are four possibilities for the fields of $a$ and $a^\prime$
incident onto the beam splitter ($BS1$), which gives the output
(In the following, only the cat part for a component of the mixed
state is shown to describe the action of the apparatuses)
\begin{eqnarray}
\label{eq:tr1}
&&|\alpha\rangle_a|\alpha\rangle_{a^\prime}
\longrightarrow|\sqrt{2}\alpha\rangle_f|0\rangle_{f^\prime},\\
&&|\alpha\rangle_a|-\alpha\rangle_{a^\prime}
\longrightarrow|0\rangle_f|\sqrt{2}\alpha\rangle_{f^\prime},\label{eq:tr2}\\
&&|-\alpha\rangle_a|\alpha\rangle_{a^\prime}
\longrightarrow|0\rangle_f|-\sqrt{2}\alpha\rangle_{f^\prime},\label{eq:tr3}\\
&&|-\alpha\rangle_a|-\alpha\rangle_{a^\prime}
\longrightarrow|-\sqrt{2}\alpha\rangle_f|0\rangle_{f^\prime}\label{eq:tr4}.
\end{eqnarray}
In the boxed apparatus P1, Alice checks if modes $a$ and $a^\prime$
were in the same state by counting photons at the photodetectors $A1$
and $A2$. If both modes $a$ and $a^\prime$ are in $|\alpha\rangle$
or $|-\alpha\rangle$, $f^\prime$ is in the vacuum, in which case the
output field of the beam splitter $BS2$ is
$|\alpha,-\alpha\rangle_{t1, t2}$.  Otherwise, the output field is
either $|2\alpha,0\rangle_{t1, t2}$ or $|0,2\alpha\rangle_{t1, t2}$.
When both the photodetectors $A1$ and $A2$ register any photon(s),
Alice and Bob are sure that the two modes $a$ and $a^\prime$ were in
the same state but when either $A1$ or $A2$ does not resister a
photon, $a$ and $a^\prime$ were likely in different states. 
The remaining pair is selected only when Alice and Bob's
all four detectors click together. Of course,
there is a probability not to resister a photon even though the two
modes were in the same state, which is due to the nonzero overlap of
$|\langle0|\sqrt{2}\alpha\rangle|^2$. Note that inefficiency of the
detectors does not degrade the the quality of the distilled
entangled coherent states
but decreases the success probability.

It can be simply shown that the second and third terms of
Eq.~(\ref{eq:rho*2}) are always discarded by the action of P1 and
Bob's apparatus same as P1.
For example, at the output ports of $BS1$ and Bob's beam splitter
corresponding to $BS1$, $|\Phi_+\rangle_{ab}|\Psi_+\rangle_{a^\prime
  b^\prime}$ becomes
\begin{widetext}  
\begin{equation}
|\Phi_+\rangle_{ab}|\Psi_+\rangle_{a^\prime b^\prime}  
 \longrightarrow N_+^2\big(|\sqrt{2}\alpha,0,0,\sqrt{2}\alpha\rangle
+|0,\sqrt{2}\alpha,\sqrt{2}\alpha,0\rangle
+|0,-\sqrt{2}\alpha,-\sqrt{2}\alpha,0\rangle
+|-\sqrt{2}\alpha,0,0,-\sqrt{2}\alpha\rangle\big)_{fgf^\prime  g^\prime},
\label{bs2}
\end{equation}
\end{widetext}  
where $g$ and $g^\prime$ are the output field modes from Bob's beam
splitter corresponding to $BS1$.
The fields of modes $f^\prime$ and $g^\prime$ can never be in
$|0\rangle$ at the same time; at least, one of the four detectors of
Alice and Bob must not click.  The third term of Eq.~(\ref{eq:rho*2}) can be shown to lead
to the same result by the same analysis.

For the cases of the first and fourth terms in Eq.~(\ref{eq:rho*2}), all four detectors
may register photon(s).
After the beam splitter $BS1$, the ket of
$(|\Phi_-\rangle\langle\Phi_-|)_{ab}\otimes
(|\Phi_-\rangle\langle\Phi_-|)_{a^\prime b^\prime}$ 
 of Eq.~(\ref{eq:rho*2}) becomes
\begin{equation}
\label{+++}
|\Phi_-\rangle_{ab}|\Phi_-\rangle_{a^\prime b^\prime} 
\longrightarrow
|\Phi^\prime_+\rangle_{fg}|0,0\rangle
_{f^\prime g^\prime}-|0,0\rangle_{fg}
|\Phi^\prime_+\rangle_{f^\prime g^\prime},
\end{equation}
where $|\Phi^\prime_+\rangle
=N_+^\prime(|\sqrt{2}\alpha,\sqrt{2}\alpha\rangle
+|-\sqrt{2}\alpha,-\sqrt{2}\alpha\rangle)$
with the normalization factor $N_+^\prime$.  The normalization factor
in the right hand side of Eq.~(\ref{+++}) is omitted.  The first term
is reduced to $(|\Phi^\prime_+\rangle\langle\Phi^\prime_+|)_{fg}$ after
$(|0,0\rangle\langle0,0|)_{f^\prime g^\prime}$ is measured out by Alice
and Bob's P1's.  Similarly, the fourth term of Eq.~(\ref{eq:rho*2})
yields $(|\Psi^\prime_+\rangle\langle\Psi^\prime_+|)_{fg}$, where
$|\Psi^\prime_+\rangle$ is defined in the same way as
$|\Phi_+^\prime\rangle$, after $(|0,0\rangle\langle0,0|)_{f^\prime g^\prime}$
is measured.  Thus the density matrix for the
field of modes $f$ and $g$ conditioned on simultaneous measurement of
photons at all four photodetectors is
\begin{equation}
\label{mid}
\rho_{fg}=F^\prime|\Phi_+^\prime\rangle\langle
\Phi_+^\prime|+(1-F^\prime)|\Psi_+^\prime\rangle\langle\Psi_+^\prime|,
\end{equation}
where 
$F^\prime=F^2/\{F^2+(1-F)^2\}$,
and $F^\prime$ is always larger than $F$ for
any $F>1/2$.
By reiterating this process, Alice and Bob can
distill some maximally entangled states $|\Phi_+\rangle$
of a large amplitude
asymptotically.  
Of course, a sufficiently large ensemble 
and initial fidelity $F>1/2$
are required for successful purification \cite{Bennett96}.

We now apply our scheme to a realistic example in a dissipative environment. 
When the entangled coherent channel $|\Phi_-\rangle$ is embedded in a
vacuum, the channel decoheres and becomes a mixed state of
its density operator $\rho_{ab}(\tau)$, where $\tau$ stands for the
decoherence time.  By
solving the master equation \cite{Phoenix}
\begin{equation}
\begin{aligned}
&{\partial \rho \over \partial \tau}=\hat{J}\rho +\hat{L}\rho~;~~\\
&\hat{J}\rho=\gamma \sum_i a_i\rho a_i^\dag,~~
\hat{L}\rho=-{\gamma \over 2}\sum_i(a_i^\dag a_i\rho +\rho a_i^\dag a_i)
\label{master-eq}
\end{aligned}
\end{equation}
where $\gamma$ is the energy decay rate, the mixed state
$\rho_{ab}(\tau)$ can be straightforwardly obtained as
\begin{eqnarray}
&&\rho_{ab}(\tau)={\widetilde N}(\tau)
\Big\{|t\alpha, t\alpha \rangle\langle
 t\alpha, t\alpha|+|-t\alpha,
-t\alpha \rangle\langle -t\alpha, -t\alpha| \nonumber \\
&&~~~~~~~~-\Gamma(|t\alpha, t\alpha \rangle\langle -t\alpha,
-t\alpha|+|-t\alpha, -t\alpha\rangle\langle
t\alpha, t\alpha|)\Big\},\nonumber \\
\end{eqnarray}
where $|\pm t\alpha,\pm t\alpha\rangle=|\pm t\alpha\rangle_a|\pm
t\alpha\rangle_b$, $t=e^{-\gamma \tau/2}$,
$\Gamma=\exp[-4(1-t^2)|\alpha|^2]$, and ${\widetilde N}(\tau)$ is the normalization
factor.
The decohered state $\rho_{ab}(\tau)$ may be represented by the
dynamic Bell-cat states defined as follows:
\begin{eqnarray}
\label{eq:dqbs1}
&&|{\widetilde \Phi}_\pm\rangle_{ab}
={\widetilde N}_\pm(|t\alpha\rangle_a|t\alpha\rangle_b
\pm|-t\alpha\rangle_a|-t\alpha\rangle_b),\\
\label{eq:dqbs2}
&&|{\widetilde \Psi}_\pm\rangle_{ab}={\widetilde
 N}_\pm(|t\alpha\rangle_a|-t\alpha\rangle_b
 \pm|-t\alpha\rangle_a|t\alpha\rangle_b),
\end{eqnarray}
where ${\widetilde N}_\pm=\{2(1\pm e^{-4t^2|\alpha|^2})\}^{-1/2}$.
The decohered state is then
\begin{eqnarray}
\label{decs}
&&\rho_{ab}(\tau)={\widetilde N}(\tau)\big\{\frac{(1+\Gamma)}
{{\widetilde
  N}_-^2}|{\widetilde \Phi}_-\rangle\langle
  {\widetilde \Phi}_-|+\frac{(1-\Gamma)}{{\widetilde
  N}_-^2}|{\widetilde \Phi}_+\rangle\langle
  {\widetilde \Phi}_+|\big\} \nonumber   \\
\label{decay}
&&~~~~~~~~\equiv F(\tau)|{\widetilde \Phi}_-\rangle
\langle{\widetilde \Phi}_-|+(1-F(\tau))|{\widetilde \Phi}_+\rangle\langle{\widetilde \Phi}_+|
\end{eqnarray}
where, regardless of the decay time $\tau$, $|{\widetilde \Phi}_-\rangle$ is 
maximally entangled and $|{\widetilde \Phi}_-\rangle$ and
$|{\widetilde \Phi}_+\rangle$
are orthogonal to each other.
 The decohered state (\ref{decay}) 
is not in the same form as Eq.~(\ref{ensemble}) so that we need some
bilateral unitary transformations before the purification scheme is applied. 
A Hadamard gate $H$ for coherent-state qubits
can be used to transform the state (\ref{decs}) into a distillable form
\begin{equation}
\label{ts2}
H_a H_b\rho_{ab}(\tau)H_b^\dagger H_a^\dagger \nonumber \\
=
F(\tau)|{\widetilde \Psi}_+\rangle\langle
{\widetilde \Psi}_+|+(1-F(\tau))|{\widetilde \Phi}_+\rangle
\langle{\widetilde \Phi}_+|,
\end{equation}
which is now in the same form as Eq.~(\ref{ensemble}).

The ensemble of state (\ref{decay}) can be purified successfully only when $F(\tau)$ is
larger than 1/2. Because $F(\tau)$ is obtained as
\begin{equation}
F(\tau)=\frac{N_+^2(1+\Gamma)}{N_+^2(1+\Gamma)-N_-^2(1-\Gamma)},
\end{equation}
it is found that purification is successful
when the decoherence time $\gamma\tau<\ln 2$ regardless of $\alpha$.  This result is in agreement with
the decay time until which teleportation can be successfully
performed via an entangled coherent state shown in Ref. \cite{JKL01}.

\section{Production of Schr\"odinger Cat States}

A key requirement of quantum information processing with cat states
is the generation of cat states in free-propagating optical fields.
This has been known to be extremely demanding 
using current technology because
strong nonlinearity \cite{Yurke} or
precise photon counting measurements \cite{Song,Dakna} are 
necessarily required. However, 
very recently, there has been remarkable progress
which may enables one to generate free propagating cat states
 without strong nonlinearity or photon counting measurements.  
For example, it was  shown that
cat states of reasonably large amplitudes 
can be produced with simple
linear optics elements and single photons \cite{Lund04}. 
Relatively small
nonlinearity  was shown to be still useful
with conditioning homodyne detection \cite{Jeong04}
or with a single photon interacting with a coherent state
\cite{Nemoto}
to generate cat states.
It was shown that a deterministic cat-state source
can be obtained using a single-atom cavity \cite{WD}.
A recent experimental progress \cite{Wenger04}
could be directly improved by
the cat-amplification scheme in Ref.~\cite{Lund04}
to generate a cat state of a larger amplitude and higher fidelity.
The above proposals have now brought
the generation of
free-propagating cat states of $\alpha\approx 2$ 
within reach of current technology. 
Electromagnetically induced transparency (EIT) has also been studied
as a method to obtain a giant Kerr nonlinearity \cite{Schmidt96,Hau99,Lukin00},
and there has been an improved suggestion to generate cat states
with it \cite{Mauro}.
In what follows, some of these suggestions
 will be briefly covered.

\subsection{Schemes using linear optics elements}

\begin{figure}
\centerline{\scalebox{0.5}{\includegraphics{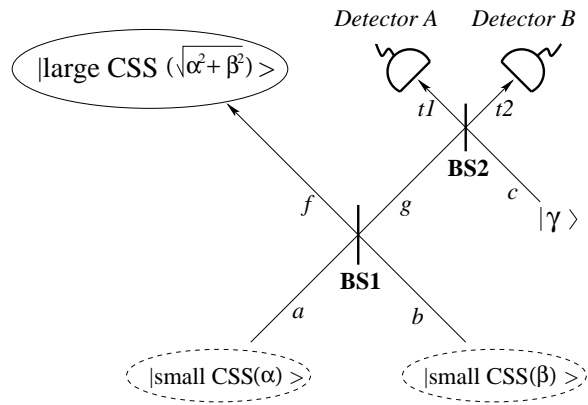}}}
\caption{
A schematic of the non-deterministic cat-amplification process.
See text for details.}
\label{fig-1uqc}
\end{figure}

Since it is extremely hard  to generate cat states
using $\chi^{(3)}$ nonlinearity,    
some alternative methods have
been studied 
based upon conditional measurements
\cite{Song,Dakna}. A crucial drawback of these schemes
is that a highly efficient photon counting measurement,
which is extremely demanding in current technology, is necessary. 
However, it was shown 
recently that a free-propagating optical cat state
can be generated with a single photon
source and simple optical operations without efficient photon detection
\cite{Lund04}.
This suggestion contains two main points: 
\begin{itemize}
\item An arbitrarily large cat state can be produced
out of arbitrarily small cat states using the simple experimental set-up 
depicted in Fig.~\ref{fig-1uqc}.
\item A small odd cat state with $\alpha \leq 1.2$ is very 
well approximated by a squeezed single photon, 
$S(s)|1\rangle$, where
$S(s)$ is the squeezing operator
with the squeezing parameter $s$
 and $|1\rangle$ is the single-photon state.
\end{itemize}
Firstly, the cat-amplification 
process summarized as follows.
Suppose two small cat states, 
$|{\rm cat}_\varphi(\alpha)\rangle$ and $|{\rm cat}_\phi(\beta)\rangle$,
with amplitudes $\alpha$ and $\beta$. 
The reflectivity $r$ and transmitivity $t$ of BS1
 are set to $r=\beta/\sqrt{\alpha^2+\beta^2}$ and 
$t=\alpha/\sqrt{\alpha^2+\beta^2}$, where 
the action of the beam splitter is represented by
${\hat B}_{a,b}(r,t)|\alpha\rangle_a|\beta\rangle_b=
|t\alpha+r\beta\rangle_f|-r\alpha+t\beta\rangle_g$.
The other beam splitter BS2 is a 50:50 beam splitter ($r=t=1/\sqrt{2}$)
regardless of the conditions and
the amplitude $\gamma$ of the auxiliary coherent field is 
determined as $\gamma=2\alpha\beta/\sqrt{\alpha^2+\beta^2}$.
The resulting state for mode $f$ then becomes 
$|{\rm cat}_{\varphi+\phi}({\cal A})\rangle\propto|{\cal A}\rangle
+e^{i(\varphi+\phi)}|-{\cal A}\rangle$,
whose coherent amplitude ${\cal A}=\sqrt{\alpha^2+\beta^2}$
 is larger than both $\alpha$ and $\beta$.
The relative phase of the resulting cat state
is the sum of the relative phases of the input cat states.
The success probability $P_{\varphi,\phi}(\alpha,\beta)$ for a single iteration 
is 
\begin{equation}
P_{\varphi,\phi}(\alpha,\beta)
=\frac{(1-e^{-\frac{2\alpha^2\beta^2}{\alpha^2+\beta^2}})^2
[1+\cos(\varphi+\phi)e^{-2(\alpha^2+\beta^2)}]}
{2(1+\cos\varphi e^{-2\alpha^2})(1+\cos\phi e^{-2\beta^2})},
\nonumber
\end{equation}
which approaches 1/2 as the amplitudes of initial cat states
becomes large. Note that the probabilities depend on the type of cat states
(odd or even) used. The effect of detector inefficiency is just to decrease
this success probability. 

\begin{figure}
\centerline{\scalebox{0.5}{\includegraphics{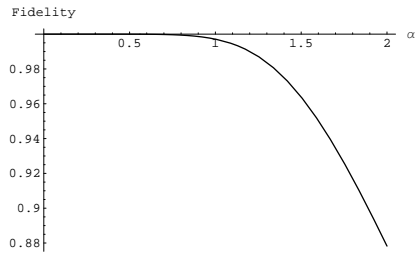}}}
\caption{The fidelity between an odd cat state and squeezed single photon.
The odd cat state is extremely well approximated by the
squeezed single photon for a small coherent amplitude, $\alpha\leq1.2$. }
\label{fid-nopa}
\end{figure}

Secondly, the fidelity between the squeezed single photon and the cat state
is
\begin{equation}
F(s,\alpha) 
=\frac{2\alpha^2\exp[\alpha^2(\tanh s-1)]}
{(\cosh s)^3(1-\exp[-2\alpha^2])},
\nonumber
\end{equation}
where $\alpha$ is the
amplitude of the cat state.
Fig.~\ref{fid-nopa} shows the maximized fidelity on the y-axis plotted
against a range of possible values for $\alpha$ for the desired odd cat state.
Some example values are:
 $F=0.99999$ for amplitude $\alpha=1/2$,
  $F=0.9998$ for $\alpha=1/\sqrt{2}$, 
and  $F=0.997$ for $\alpha=1$, where the maximizing squeezing parameters are 
$s=0.083$, $s=0.164$, and $s=0.313$ respectively. 
The fidelity approaches unity
for $\alpha$ very close to zero while the fidelity tends to zero
as $\alpha$ increases.

The two points explained above can be efficiently combined. If one uses
squeezed single photons as small cat states, a cat state of
high fidelity ($F>0.99$) can be obtained  up to $\alpha=2.5$. 
Another interesting aspect of this process is that
it is somewhat resilient to the photon production inefficiency because
its first iteration purifies the mixed cat states while amplifying them.
For example, if the inefficiency of the single photon source is
about 40\%, the fidelity of the initial cat,
which is a mixture with a squeezed vacuum,
is $F\approx0.60$ but it will become $F\approx0.89$ by the first iteration.

It is also important
to note that there is an alternative method to obtain a squeezed single
photon even without a single photon source. 
An interesting observation is that a squeezed single photon can be
obtained by subtracting a photon from a squeezed vacuum.
This can be shown by applying the annihilation operator to a squeezed
single photon as 
$\hat a S(s)|0\rangle=\cosh s S(s)|1\rangle$.
In a recent experiment \cite{Wenger04},
the single photon subtraction was approximated
by a beam splitter of low reflectivity and a single photon detector.
Such an experiment could be immediately linked to our suggestion to
experimentally generate a larger cat state. One can then generate 
a cat state of $\alpha>2$ using our scheme without a single photon source.

Since this scheme uses at least two beam splitters to mix propagating fields,
good mode matching is required to obtain a cat state of high quality. 
Highly efficient mode matching of a single photon
from parametric down conversion and a weak coherent state from an attenuated laser beam 
at a beam splitter has been experimentally demonstrated
using optical fibers \cite{Pitt}. Such techniques could be employed for
the implementation of our scheme.
The success probability will rapidly drop down
and the required resources will exponentially 
increase as the number of steps increases.
However, if quantum optical memory is available,
one can considerably boost up the success probability by holding
the resulting states for every step \cite{JeongUn}.

\subsection{Schemes using cavity quantum electrodynamics}

Cavity quantum electrodynamics (QED) has been studied to
enhance nonlinear effects to generate macroscopic superpositions
\cite{Tu}.
Some success has been reported in creating such superposition states within
high Q cavities in the microwave \cite{MB} and optical \cite{Mon} domains.
Simplified versions of cavity QED schemes have
been developed for deterministic generation of cat states in a cavity
\cite{R04}.
While this method is relatively effective to generate cat states in cavity,
most of the schemes suggested for quantum
information processing with coherent states
require {\it free propagating} cat states.

Recently, a method was proposed to generated free propagating cat states
by using cavity-assisted interactions of coherent optical pulses 
\cite{WD}.
This suggestion employs an atom of three relevant levels trapped in an optical
cavity with a coherent-state pulse $|\alpha\rangle$ incident onto the cavity.
One of the atomic level, $|e\rangle$, is its excited state, and the
other two levels, $|g_0\rangle$ and $|g_1\rangle$, are
levels in the ground state with different hyperfine spins.
The transition from $|g_1\rangle$ to $|e\rangle$ is resonantly coupled
to a cavity mode while $|g_0\rangle$ is decoupled from the cavity mode.
In such a preparation, if the trapped atom was prepared in state $|g_0\rangle$,
the input field becomes $|-\alpha\rangle$ after a resonant reflection
as the input pulse is resonant with the bare cavity mode.
On the other hand, if the atom was prepared in state $|g_1\rangle$,
it remains $|\alpha\rangle$ due to a strong atom-cavity coupling.
Therefore, if the trapped atom was prepared in a superposition state 
such as $(|g_0\rangle+|g_1\rangle)/\sqrt{2}$,
the reflected field becomes an entangled state
$(|g_0\rangle|\alpha\rangle+|g_1\rangle|-\alpha\rangle)/\sqrt{2}$,
which can be projected to a single mode cat state by a measurement
on a superposed basis $|g_0\rangle\pm|g_1\rangle$.
An advantage of this scheme is weak dependence on dipole coupling,
but wave front distortion
due to difference between resonant and non-resonant interactions
could be a problem in a real experiment.
This work \cite{WD} concludes that a cat state with a quite large amplitude
($\alpha\approx3.4$) could be generated in this way with a
90\% fidelity using current technology.

\subsection{Schemes using weak nonlinearity}

There has been a suggestion to use relatively weak nonlinearity
with beam splitting with a vacuum  and 
conditioning by homodyne detection to generate
cat states \cite{Jeong04}.
As beam splitting with a vacuum and homodyne measurement 
can be highly efficient
 in quantum optics laboratories, this shows that relatively weak nonlinearity can still
be useful to generate cat states.

The Hamiltonian of a single-mode Kerr
nonlinear medium is 
${\cal H}_{NL}=\omega a^\dag a+\lambda(a^\dag a)^2$, 
where $a$ and $a^\dag$ are annihilation and creation operators,
$\omega$ is the energy level splitting for the
harmonic-oscillator part of the Hamiltonian and $\lambda$ is the
strength of the Kerr nonlinearity \cite{Yurke}. Under the influence of the
nonlinear interaction the initial coherent state $|\alpha\rangle$
evolves to the following state at
time 
$\tau=\pi/\lambda N$
\cite{LKLB}:
\begin{equation}
\label{c2}
|\psi_N\rangle=\sum_{n=1}^N C_{n,N}|-\alpha
 e^{2in\pi/N}\rangle,
\end{equation}
where 
\begin{equation}
C_{n,N}=
\frac{1}{N}\sum_{k=0}^{N-1}(-1)^k
\exp[-\frac{i\pi k}{N}(2n-k)].
\end{equation}
 The length $L$ of the nonlinear cell
corresponding to $\tau$ is $L=v\pi/2\lambda N$, where $v$ is the
velocity of light.  For $N=2$,
 we obtain a desired cat state of
relative phase $\varphi=\pi/2$. 
We again emphasize the nonlinear coupling $\lambda$ is typically
very small such that $N=2$ cannot be obtained in a length limit where
the decoherence effect can be neglected.

 If
$\lambda$ is not as large as required to generate the cat state,
the state (\ref{c2}) with $N>2$ may be obtained
by choosing an appropriate interaction time.
  From the state (\ref{c2}),
it is required to remove all the other coherent component states
except two coherent states of a $\pi$ phase difference.  First,
it is assumed that the state (\ref{c2}) is incident on a 50-50 beam
splitter with the vacuum onto the other input of the beam
splitter. The initial coherent
amplitude $\alpha_i$ is taken to be real for simplicity. The
state (\ref{c2}) with initial amplitude $\alpha_i$ after passing
through the beam splitter becomes
\begin{equation}
\label{cbv}
|\psi_N\rangle=\sum_{n=1}^N C_{n,N}|-\alpha_i e^{2in\pi/N}/\sqrt{2}\rangle
|-\alpha_i e^{2in\pi/N}/\sqrt{2}\rangle,
\end{equation}
where all $|C_{n,N}|$'s have the same value. The real part
of the coherent amplitude in the state (\ref{cbv}) is then
measured by homodyne detection
in order to produce the cat state in the other path.
 By the measurement result, the
state is reduced to
\begin{eqnarray}
\label{cm} |\psi^{(1)}_N\rangle= \sum_{n=1}^N
C^{(1)}_{n,N}(\alpha_i) |-\alpha_i e^{2in\pi/N}/\sqrt{2}\rangle,
\end{eqnarray}
where $C^{(1)}_{n,N}(\alpha_i) = {\cal N}_\psi \sum_{n=1}^N
C_{n,N } \langle X|-\alpha_i e^{2in\pi/N}/\sqrt{2}\rangle$ with
${\cal N}_\psi$ the normalization factor and $|X\rangle$ the
eigenstate of ${\hat X}=(a+a^\dagger)/\sqrt{2}$.
 After the
homodyne measurement, the state is selected when the measurement
result is in certain values.  If coefficients
$|C^{(1)}_{N/2,N}(\alpha_i)|$ and $|C^{(1)}_{N,N}(\alpha_i)|$ in
Eq.  (\ref{cm}) have the same nonzero value and all the other
$|C^{(1)}_{n,N}(\alpha_i)|$'s are zero, then the state becomes a
desired cat state. Suppose $N=4k$ where $k$ is a positive integer number.
If $X=0$ is measured in this case, the
coefficients $|C^{(1)}_{n,N}(\alpha_i)|$'s will be the largest
when $n=N/4$ and $n=3N/4$, and become smaller as $n$ is far from these
two points. The coefficients can be close to zero for all the
other $n$'s for an appropriately large $\alpha_i$ so that the
resulting state may become a cat state of high fidelity. 
Using this technique, 
 one may observe a conspicuous
signature of a cat state even
with a 1/100 times weaker nonliearity
compared with the currently required level \cite{Jeong04}.
In particular, this approach can be useful to produce a cat state
with a significantly large amplitude such as $\alpha\geq 10$.

Another scheme \cite{Nemoto} proposed for linear optics quantum
computation \cite{KLM} uses
weak cross-Kerr nonlinearity of the interaction Hamiltonian 
$H=\hbar \chi a_1^\dagger a_1 a_2^\dagger a_2$
to generate a cat state. The 
interaction between a coherent state, $|\alpha\rangle_2$, and
a single-photon qubit, e.g.,
$|\psi\rangle_1=(|0\rangle_1+|1\rangle_1)/\sqrt{2}$, is described as
\begin{eqnarray}
U_K|\psi\rangle_1|\alpha\rangle_2&=&e^{i H_{K}t/\hbar}\frac{1}{\sqrt{2}}
(|0\rangle_1+|1\rangle_1)|\alpha\rangle_2\\
&=&\frac{1}{\sqrt{2}}(|0\rangle_1|\alpha\rangle_2+|1\rangle_1|\alpha e^{i\theta}\rangle_2),
\end{eqnarray} 
where
$|0\rangle$ ($|1\rangle$) is the vacuum (single-photon) state,
$\alpha$ is the amplitude of the coherent state,
and $\theta=\chi t$ with the interaction time $t$.
If $\theta$ is $\pi$ and one measures out mode 1 on a superposed basis 
$(|0\rangle\pm|1\rangle)/\sqrt{2}$, a macroscopic superposition state
(so-called Schr\"odinger cat state),
$(|\alpha\rangle\pm|-\alpha\rangle)/\sqrt{2}$.
Using dual rail logic instead of the superposition between
the single photon and the vacuum, 
the measurement on the superposed basis can be simply realized with
a beam splitter and two photodetectors \cite{Gerry}.
It is again extremely difficult to obtain $\theta=\pi$ using
currently available nonlinear media.
However, simply by increasing the amplitude $\alpha$,
one can gain an arbitrarily large separation between
$\alpha$ and $\alpha e^{i \theta}$ within
an arbitrarily short interaction time.
It is possible to transform
the state of the form of  $|\alpha\rangle\pm|\alpha^{i\theta}\rangle$
 to the symmetric form of
$|\alpha^\prime\rangle\pm|-\alpha^\prime\rangle$ by 
the displacement operation which can be simply performed using
a beam splitter with the transmission coefficient close to one and 
a strong coherent state being injected into the other input port.   
Therefore, weak cross-Kerr nonlinearity can also
be useful to generate a
cat state with a single photon, strong coherent states,
beam splitters and two photodetectors.
Remarkably, it can be shown that this approach
can also reduce decoherence effects
by increasing the initial amplitude $\alpha$,
which is also true for Ref.~\cite{Jeong04}.
It should also be noted that the detectors
and the single photon source
in Ref.~\cite{Gerry} 
which can be directly combined with Ref.~\cite{Nemoto}
do not have to be efficient
for a conditional generation of a cat state
 because these factors only degrade
the success probability to be less than unity.

\section{Conclusion}

We have discussed quantum processing tasks using qubits based on coherent states and Schr\"odinger cat 
states as resources. We have shown that a universal set of processing tasks can be achieved using only 
linear optics, feedforward and photon counting. This is a similar result to that of Knill Laflamme and 
Milburn for single photon qubits \cite{KLM}. However, far fewer operations per gate are needed in the 
coherent state scheme and shortcuts are available for certain tasks. On the other hand these advantages 
are not useful unless a good way of producing cat states can be found. Thus we have spent some time 
discussing various proposals, both linear and non-linear, for producing cat states. We believe the near 
term prospects for demonstrating small travelling-wave cat states and basic processing tasks based on 
them are good. Whether coherent state qubits or single photon qubits will prove better for larger scale 
quantum optical processing in the long run remains an open question.

\end{document}